\documentclass{sig-alternate}
\usepackage{verbatim}
\usepackage{times}
\usepackage{enumitem}
\usepackage{hyperref}
\usepackage{graphicx}
\hypersetup{
	colorlinks,
	citecolor=blue,
	filecolor=black,
	linkcolor=blue,
	urlcolor=blue,   
}

\newtheorem{theorem}{\textbf{Strategy}}

\makeatletter
\let\@copyrightspace\relax
\makeatother
\begin{document}

\title{Learning Optimal Card Ranking from Query Reformulation}

\numberofauthors{1}
\author{Liangjie Hong, \, Yue Shi,  \, Suju Rajan\\
  \affaddr{Personalization Sciences, Yahoo Research, Sunnyvale CA, USA}\\
  \email{\{liangjie,yueshi,suju\}@yahoo-inc.com}
}

\maketitle
\begin{abstract}
Mobile search has recently been shown to be the major contributor to the growing search market. The key difference between mobile search and desktop search is that information presentation is limited to the screen space of the mobile device. Thus, major search engines have adopted a new type of search result presentation, known as \textit{information cards}, in which each card presents summarized results from one domain/vertical, for a given query, to augment the standard blue-links search results. While it has been widely acknowledged that information cards are particularly suited to mobile user experience, it is also challenging to optimize such result sets. Typically, user engagement metrics like \textit{query reformulation} are based on whole ranked list of cards for each query and most traditional learning to rank algorithms require per-item relevance labels. In this paper, we investigate the possibility of interpreting query reformulation into effective relevance labels for query-card pairs. We inherit the concept of conventional learning-to-rank, and propose pointwise, pairwise and listwise interpretations for query reformulation. In addition, we propose a learning-to-label strategy that learns the contribution of each card, with respect to a query, where such contributions can be used as labels for training card ranking models. We utilize a state-of-the-art ranking model and demonstrate the effectiveness of proposed mechanisms on a large-scale mobile data from a major search engine, showing that models trained from labels derived from user engagement can significantly outperform ones trained from human judgment labels.
\end{abstract}

\vspace{1mm}
\noindent
{\bf Categories and Subject Descriptors:} H.3.5 {[Information Storage and Retrieval]}: {Online Information Services}

\vspace{1mm}
\noindent
{\bf General Terms:} Design, Theory, Experimentation

\vspace{1mm}

\section{Introduction}\label{sec:introduction}
\begin{figure}[t]
	\centering
	\includegraphics[scale=0.25]{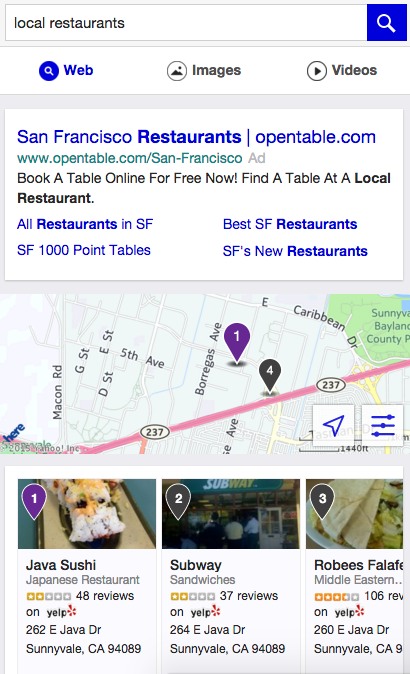}
	\includegraphics[scale=0.25]{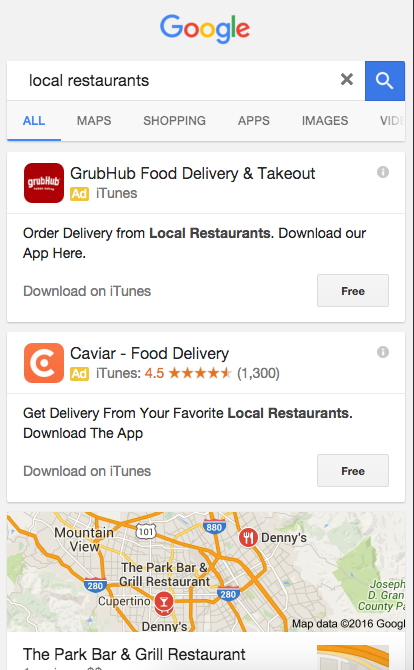}
	\caption{\textbf{A snapshot of search results from two major mobile search engines for the query ``{\tt Local Restaurants}'' where enriched restaurant information is shown in a {\tt LocalCard}.}}
	\label{fig:cards}
\end{figure}

Mobile search has recently been reported to be the major contributor to the search market\footnote{www.businessinsider.com/google-search-traffic-mobile-passes-desktop-2015-5}. The key difference of mobile search from traditional desktop search lies in the fact that information presentation is constrained to the screen space of a mobile device. For this reason, major search engines have adopted a new type of search result presentation, known as \textit{information cards}\footnote{www.google.com/landing/now/\#cards}, in which, for a given query, each card presents summarized results from one domain/vertical. For instance, when a user types ``{\tt Local Restaurants}'' on a mobile device, modern search engines can directly pull out maps and relevant restaurant information organized in a clean and concise way, shown in Figure \ref{fig:cards}. Another example is that, given the query ``{\tt Brad Pitt}'', a {\tt PersonCard} can be triggered to present relevant information about this person, e.g., bio, photos and recent movie show. In addition, a {\tt NewsCard} can be also triggered with the latest news articles concering the actor. The core concept is that, with one or a small number of cards, a user's information needs can be satisfied directly, without scrolling down to the regular or conventional web search results. It has been widely recognized that information cards are well-suited to mobile devices and have greatly improved the mobile search experience\footnote{blog.intercom.io/why-cards-are-the-future-of-the-web}.
While information cards can effectively augment traditional web results, the definition of \textit{a successful search} or whether a user is satisfied with a particular ranking of cards, with respect to a query, is becoming more difficult to answer. In the conventional notion of relevance, each search result is judged on whether it is relevant to a query using a graded scale. Such judgments are static and unique for a query-search result pair. Nevertheless, these judgments laid the foundation for a number of search metrics like Normalized-Discounted-Cumulative-Gain ({\tt NDCG})~\cite{Jaervelin2002} or Expected-Reciprocal-Rank ({\tt ERR})~\cite{Chapelle2009}. In recent years, researchers also found that click-based implicit feedback~\cite{Joachims2002,Joachims2007} is a very important signal for relevance. However, for the problem of card ranking, the above feedback mechanisms are lacking. For instance, the relevance between a card and a query can be temporal and the relative ordering of such relevance among different cards might be subtle and can be very different for different users. Additionally, click-based signals may not be available. For example, a \textit{Question-to-Answer} ({\tt Q2ACard}) card could be relevant to a query, if the user sees it and finds it to directly answer his/her question thereby not requiring any further action on the user's part. Therefore, it is impossible to solely depend on click-based signals as labels for card ranking. In order to mitigate these issues, search engines have started resorting to \textit{Query Reformulations}, whether a user quickly reforms his/her current query into a similar query, as one of the key online metrics to evaluate the quality of search results~\cite{Hassan2013,Huang2009,Kelly2003}. The core assumption is that if a user is satisfied with the search results, he/she is less likely to reformulate the query to search further within that query session. The same metric has also shown great impact on mobile search~\cite{Lagun2014,Shokouhi2014}. For this reason, we can expect that a good card ranking model should, for each query, rank the cards in a way that the user is less likely to reformulate his/her query. Therefore, the target of a card ranking model is ultimately to reduce the amount of query reformulations.

However, it is challenging to optimize query reformulations directly for two reasons. Firstly, a query reformulation event is not defined on a per query-result level and it is usually realized when a user inputs two similar (a predefined threshold is used to determine whether two queries are similar or not) queries consecutively. Therefore, how to interpret such an observation of query reformulation from the query level to the level of each query-card pair can be seen as the core of many research challenges. Second reason is that, many cards can be relevant to a query even without any click (e.g., {\tt Q2ACard} and {\tt WeatherCard}). Thus, conventional labeling strategies based on user's implicit feedback, such as clicks and skips on the search results~\cite{Joachims2002,Joachims2007}, are not applicable to cards. One may consider that we can pre-define, for a given card, the types of user interaction, as positive/negative labels, e.g., taking the interaction of ``viewing with no click'' as the positive label for {\tt Q2ACard}. However, this strategy cannot scale in practice, where we could have a large number of cards, requiring tremendous effort to predefine all types of user interaction for each card.

In this work, we specifically investigate the possibility of interpreting query reformulations into effective relevance labels for training card ranking models. On a high level, a card ranking model shall take features extracted from queries, query-card pairs and lists of cards as input and output a ranked list of cards for the query. Although remarkable feature engineering is required, we emphasize that the key to training an effective ranking model lies in the quality of labels, which provides the learning basis of the relevance between queries and cards. In this paper, we aim to propose and discuss several strategies to derive labels for card ranking from query reformulations. Thus, card ranking models can be trained to optimize such metrics in online systems. The key research question we attempt to address in this paper is:
\begin{itemize}
\item \textit{How to learn from query reformulation for labeling query-card pairs?}
\end{itemize}
We inherit the concept of conventional learning-to-rank, and propose pointwise, pairwise and listwise interpretations for query reformulations. In addition, we propose a \textit{learning-to-label} strategy that learns the rewards for the cards shown to each query. The rewards can be further used as labels for training card ranking models. Note that our focus of the paper is not on the choice of ranking models. Throughout the paper, we will use Gradient Boosted Trees ({\tt GBT}) as the model~\cite{Friedman2001,Zheng2008}, which has been shown to be an effective ranking model for web search, including the {\tt KDD CUP} of Learning to Rank competition~\cite{Burges2011LTR}.

Our main contributions in this paper are summarized as follows:
\begin{enumerate}
	\item We systematically analyze online users' behaviors with respect to query reformulations and demonstrate that it is difficult to optimize such user engagement metric through conventional human-judgments-based learning to rank procedure.
	\item We propose several strategies to derive labels from query reformulations and provide a guideline to train ranking models based on them.
	\item We compare these mechanisms on a large-scale data set from a major search engine and show that ranking models trained from proposed method can significantly outperform the ones trained from human labels.
\end{enumerate}The paper is organized as follows. In \S\ref{sec:related_work}, we review the related work followed by \S\ref{sec:card_ranking}, in which we present the details of the proposed card ranking labeling strategies. In \S\ref{sec:gbdt}, we describe the ranking model used in our work. Then, in \S\ref{sec:experiment}, we elaborate our evaluation methodology and demonstrate the performance of the proposed labeling strategies. \S\ref{sec:conclusion} concludes the paper.

\section{Related Work}\label{sec:related_work}
Our work is related to research topics of heterogeneous web search, including vertical selection and federated search, and the work of deriving labels from user implicit feedback. In this section, we discuss each of these aspects, and position our work within that research spectrum.

\medskip

{\bfseries Heterogeneous Web Search}: Selection of one or a few relevant information domains has been extensively studied as the problem of \textit{vertical selection}. Vertical selection is one of the earliest research efforts to integrate heterogeneous information in specific domains into conventional web search. Note that the notion of ``vertical'' often used in desktop web search is equivalent to the notion of ``card'' in mobile search as both are presented a block of information from a particular domain in a search result page. Diaz~\cite{Diaz2009a} first studied selection models for news domain. Arguello et al.~\cite{Arguello2009a, Arguello2009} investigated the vertical selection problem on multiple domains, such as images and videos. As mentioned earlier, traditionally, research on vertical selection has focused on choosing a very few relevant domain and thus, methods and models developed so far are inherently based on binary classifications or multi-class classifications (e.g.,~\cite{Diaz2009a,Diaz2009,Arguello2009,Arguello2009a,Ponnuswami2011,Zhou2012}) where decisions are made towards each vertical to be selected or not. While it is possible to obtain a ranking from independent binary decisions, the relative ordering of verticals were not modeled explicitly in those works. Aggregated search is then established as another research topic that dedicates to building models for ranking multiple verticals. In this sense, the notion of ``aggregated search'' is technically equivalent to the notion of ``card ranking'' in mobile search. Note that the area of \textit{federated search}~\cite{Shokouhi2011} in literature is similar to aggregated search, since both of them are built by merging information from a variety of verticals. Arguello et al.~\cite{Arguello2011} made a further contribution to aggregated search by a comprehensive analysis of both feature engineering and modeling approaches. This work is closely related to our work in the sense that we also target to build effective card ranking models. However, the difference is still significant, since our focus is on deriving card ranking labels from users' reformulation events, while the early work of~\cite{Arguello2011} relied on human assessments, which, as discussed in \S\ref{sec:introduction}, is not scalable. For more related literature along this line, please refer to \cite{Shokouhi2011}.

\medskip

{\bfseries Labeling from User Feedback}: Most of previous work resorts to human experts to judge relevance labels in the problem of vertical selection and aggregated search (e.g.,~\cite{Arguello2011,Arguello2011a,Zhou2014}). Recently, a few contributions have been made to exploit implicit feedback signals from online users to derive the query-vertical (comparable to query-card) relevance. Particularly, a large body of those contributions are in the similar spirit to the fundamental work by Thorsten et al.~\cite{Joachims2002,Joachims2007a,Joachims2007} that exploits user implicit feedback, i.e., clicks and skips, for optimizing conventional web search. Ponnuswami et al.~\cite{Ponnuswami2011} has provided a method to derive labels of verticals based on user click data. A label for a particular query-vertical pair is determined by whether the vertical has received a click from the user, and the relative position of the vertical to the first web results block. Jie et al.~\cite{Jie2013} has casted the vertical ranking problem into a multi-armed bandit problem and tried to learn a regression function to predict the rewards for each vertical shown to different positions. The rewards used in their paper are defined by click-skip actions from users on each vertical. As mentioned in \S\ref{sec:introduction}, those works are inherently not applicable to the cases where the relevance between a vertical and a query cannot be measured based on users' click/skip actions. In contrast, our work is directly motivated from online metric of query reformulations, which are completely decoupled from users' click data. It is worth mentioning that one early work in conventional web search has exploited query reformulations for deriving the labels of web results~\cite{Joachims2007}. However, those papers only use query reformulations to extend the click/skip based labeling strategy, i.e., it still relies on user click data. For this reason, our work is substantially different from previous work not only because we tackle the problem of card ranking, not that of conventional web search, but also because we propose new strategies to exploit query reformulations, independent of click signals.

\medskip

{\bfseries Information Cards}: To our knowledge, there is no prior work on the problem of card ranking in mobile search. One of the latest work by Shokouhi and Guo~\cite{Shokouhi2015} was among the first to present the problem of serving cards to mobile users. However, their work and follow-up papers like \cite{Yang2016} addressed the problem under the \textit{proactive} search setting, i.e., generating card recommendations \textbf{without} queries from the user. Our work, on the other hand, address the typical search problem in which the cards are ranked in response to the user's explicit queries.

\medskip

{\bfseries Learning to Rank}: This work is related to the field of learning-to-rank ({\tt LtR}) but with significant difference. The main focus of {\tt LtR} literature is to investigate ranking models for settings where relevance labels are available. Even in the so-called \textit{listwise} case (e.g.,~\cite{Cao2007,Xia2008}) where ranking models are explicitly trained against a list of results for a particular query, the availability of relevance labels for each query-result pair is a prerequisite. Indeed, as mentioned above, the classic setting of {\tt LtR} \cite{Chapelle2011} requires human relevance judgments and these labels are provided on the query-result level, which is different from our setting. For more thorough discussion about {\tt LtR}, please refer to \cite{Liu2009}.

\medskip

Note that, we are aware of research efforts to define more fine-grained, or to some extend, better user engagement metrics, other than query reformulations for search results, such as dwell time~\cite{Kim2014} or more complex task-level satisfaction metrics~\cite{Jiang2015}. We use query reformulations in this paper as a reasonable starting point and leave extensions on deriving labels from other more advanced metrics for furture work.

\section{Labels for Card Ranking}\label{sec:card_ranking}
Our input data consists of $N$ query-page-view ({\tt QPV}) events $\mathcal{E} = \{\mathbf{q}_{1}, \cdots, \mathbf{q}_{N}\}$ where each {\tt QPV} $\mathbf{q}_{i} = (q_{i}, \pi_{q_{i}}, l^{q_{i}})$. Here, $q$ represents a query, drawn from $\mathcal{Q}=\{w_{1}, w_{2}, \cdots, w_{M}\}$ where $M$ is the number of distinct queries. The input data $\mathcal{E}$ could contain multiple occurances of the same query. In $\mathbf{q}$, $\pi_{q}$ is a ranking of a (sub)set of cards where each card $c$ is drawn from a whole set of cards $\mathcal{C} = \{c_{1},\cdots, c_{K}\}$. The ranking $\pi_{q} = \{c_{3} \succ c_{1} \succ c_{7}\}$ represents that for a query $q$, the card $c_{3}$ is ranked higher than $c_{1}$ and $c_{1}$ is ranked higher than $c_{7}$. For each $\pi_{q}$, we define $\mathcal{C}_{q}$ as the induced set of cards from the ranking. In the {\tt QPV} tuple, $l^{q}$, the label, is $+1$ if $q_{i}$ is not reformulated and $-1$ if $q_{i}$ is reformulated into another query $q_{i+1}$. Multiple queries can form a reformulation chain if $l^{q}$ is $-1$ for multiple consecutive queries. Note that, how $l^{q}$ is derived is out of the scope of this paper and we treat it as given. Roughly speaking, $l^{q_{i}}=-1$ if the next query $q_{i+1}$ is very similar to the current query $q_{i}$ and therefore, we believe that the user is not satisfied with the query $q_{i}$ and thus, make its label negative. Also, another key point is that, labels are defined on {\tt QPV}-level not on query-level. Thus, for the same query, different {\tt QPV} events might have different labels.

The goal of a card ranking model is to provide $\pi_{q}$ for each $q$ based on query features, card-level features, user features and contextual features. In this section, we argue that the key challenge in training the ranking model, however, lies in that how to define the relevance label for each query-card pair based on user feedback. As discussed in Section~\ref{sec:introduction}, the card ranking model should be optimized for reducing query reformulations. Therefore, we propose a few strategies to label each query-card pair based on users' query reformulation activity. 

Note that, we define a \textit{card} as a composite unit of multiple information widgets to serve a specific purpose or a task. In other words, even if the actual content of a card varies, two cards are considered the same \textit{type} of cards when they are serving the same specific purpose. For instance, a {\tt NewsCard} may have different news titles and links for two similar queries like ``{\tt Obama}'' and ``{\tt Obama News}'' but they are essentially the same card type. Thus, we use the notion of ``card'' to represent a particular card type throughout the paper.

\subsection{Pointwise Labeling}\label{sec:pointwise}
The first and most intuitive proposal we have is to directly translate a query-reformulation event from the query-level to the set of cards that are involved in each {\tt QPV}. Namely, we need to derive $l_{c_{k}}^{q}$ for $c_{k} \in \mathcal{C}_{q}$, the label for card $c_{k}$, based on $l^{q}$. We start the most straightforward one as below:
\begin{theorem}[Na\"{i}vely Pointwise Labeling]
Let $l_{c_{k}}^{q_{i}} = +1$ for all $c_{k} \in \mathcal{C}_{q_{i}}$ when $l^{q_{i}} = +1$. Let $l_{c_{k}}^{q_{i}} = -1$ for all $c_{k} \in \mathcal{C}_{q_{i}}$ when $l^{q_{i+1}} = +1 \wedge l^{q_{i}}=-1$. 
\end{theorem}Under this definition, we treat all query-card pairs that appeared in one {\tt QPV}, which is later reformulated, as equally negative examples, while all query-card pairs that appeared in one {\tt QPV} which is reformulated from a query but not further reformulated as equally positive examples. Note that, we only consider to derive labels for cards in the last {\tt QPV} which has the negative label while ignoring all previous ones in the chain is because users might explore their information needs in a chain of queries and query reformulations may not be equally bad in those cases. Treating all $l_{c_{k}}^{q_{i}}=-1$ for all $q_{i}$ no matter $l^{q_{i+1}}=+1/-1$ is an interesting future work. We illustrate this strategy in the following example. Suppose we have $l^{q_{2}} = +1$ and $l^{q_{1}} = -1$ and their corresponding ranked cards are shown below:
\begin{align}\label{eq:example_1}
\pi_{q_{1}}: c_1 \succ c_2 \succ c_3 \succ c_4 \nonumber \\
\pi_{q_{2}}: c_3 \succ c_2 \succ c_5 \succ c_4
\end{align}
Then, we can derive query-card labels from this reformulation activity as shown below:
\begin{align}
l^{q_{1}}_{c_{1}} = -1,  l^{q_{1}}_{c_{2}} = -1, l^{q_{1}}_{c_{3}} = -1, l^{q_{1}}_{c_{4}} = -1\nonumber \\
l^{q_{2}}_{c_{3}} = +1,  l^{q_{2}}_{c_{2}} = +1, l^{q_{2}}_{c_{5}} = +1, l^{q_{2}}_{c_{4}} = +1\nonumber
\end{align}We can see that this labeling strategy has no consideration on the relative ordering of cards. For example, an ideal model trained from these labels would have the same relevance prediction for $c_{3}$, $c_{2}$, $c_{5}$ and $c_{4}$ for the {\tt QPV} $\mathbf{q}_2$, while the relative ordering among them remains uncertain. For this reason, we take into account the rank position of each card in a query session to assign a weight to the corresponding label, leading to the following strategy.

\begin{theorem}[Discounted Pointwise Labeling]
We use $r_{k}^{q}$ denotes the rank position for the card $c_{k} \in \mathcal{C}_{q}$. A weighting function is defined as $w_k^{q} = \frac{1}{\log(1+r_{k}^{q})}$ and the label of the card $c_{k}$ is defined as $l_{c_{k}}^{q} = w_k^{q} \times l^{q}$.
\end{theorem}The main assumption behind this strategy is that the higher a card ranked in a page, the larger impact a card makes on the user's decision. For Example \ref{eq:example_1}, we may assume that the reason for the user decided to reformulate $q_1$ is mainly because he/she is not satisfied with $c_1$, the top card in $\mathbf{q}_{1}$.  Therefore, we introduce a {\tt NDCG} style discounting function to encode the impact of relative ordering of cards. Note that throughout the paper we use rank position starting from 1. For Example \ref{eq:example_1}, we could derive the following labels:
\begin{align}
l^{q_{1}}_{c_{1}} = -1.4426,  l^{q_{1}}_{c_{2}} = -0.9102, l^{q_{1}}_{c_{3}} = -0.7213, l^{q_{1}}_{c_{4}} = -0.6213\nonumber \\
l^{q_{2}}_{c_{3}} = +1.4426,  l^{q_{2}}_{c_{2}} = +0.9102, l^{q_{2}}_{c_{5}} = +0.7213, l^{q_{2}}_{c_{4}} = +0.6213\nonumber
\end{align}As we can see that, this labeling strategy not only translates positive and negative information to each cards but also keeps the relative ordering. For positive ones, it tries to encourage the ranking that matches the {\tt QPV} while for negative ones, it tries to penalize the ranking that matches the data.

Up to now, the above two labeling strategies are still based on an individual {\tt QPV}, i.e., when we define labels for {\tt QPV} $\mathbf{q}_1$, we do not take into account observations from {\tt QPV} $\mathbf{q}_2$. Therefore, the pointwise ranking model might not be effective in capturing the relative difference from two consecutive {\tt QPV}s, and thus, missing the information that affects the user's decision to reformulate. We present a labeling strategy which particularly addresses this concern. The main assumption here is that a query reformulation satisfies a user's information need because it moves relevant cards up in the list and in the meanwhile, moves irrelevant cards down in the list, or brings relevant cards into the list. Specifically, we define a movement-based pointwise labeling strategy as follows:
\begin{theorem}[Movement-based Pointwise Labeling]
	When a query $l^{q_{i}}=-1$ and $l^{q_{i+1}} = +1$, we derive $l_{c_k}^{q_{i+1}}$ for all $c_k \in \mathcal{C}_{q_{i+1}}\cup \mathcal{C}_{q_{i}}$ based on the change of card positions, as defined by a function $D(r^{q_{i}}_k, r^{q_{i+1}}_k)$. Namely, $l_{c_k}^{q_{i+1}} =D(r^{q_{i}}_k, r^{q_{i+1}}_k)$. 
\end{theorem}The key ingredient is to define the function $D$. We have five possibilities of card movement during a reformulation process, as described below:
\begin{enumerate}
	\item In the case of $r_{k}^{q_{i+1}} < r_{k}^{q_{i}}$ for $c_{k} \in \mathcal{C}_{q_{i}} \wedge c_{k} \in \mathcal{C}_{q_{i+1}}$, as $l^{q_{i}}=-1 \wedge l^{q_{i+1}} = +1$, we interpret that moving up $c_{k}$ in $q_{i+1}$ contributes to the user's satisfaction.
	\item When $r_{k}^{q_{i+1}} > r_{k}^{q_{i}}$ for $c_{k} \in \mathcal{C}_{q_{i}} \wedge c_{k} \in \mathcal{C}_{q_{i+1}}$, as $l^{q_{i}}=-1 \wedge l^{q_{i+1}} = +1$, we interpret that moving down $c_{k}$ in $q_{i+1}$ may not be relevant to the user's need.
	\item When $r_{k}^{q_{i+1}} = r_{k}^{q_{i}}$ for $c_{k} \in \mathcal{C}_{q_{i}} \wedge c_{k} \in \mathcal{C}_{q_{i+1}}$, as $l^{q_{i}}=-1 \wedge l^{q_{i+1}} = +1$, our interpretation in this case is that the card stayed in the same position has neutral impact, compared to other cards, on the user's satisfaction.
	\item If $r_{k}^{q_{i+1}} > 0$ for $c_{k} \in \mathcal{C}_{q_{i+1}}$ but $c_{k} \notin \mathcal{C}_{q_{i}}$, it is possible that the user is satisfied because of the information brought by the newly appeared card. For this reason, we interpret that the appeared card has a positive contribution to the user's satisfaction.
	\item If $r_{k}^{q_{i}} > 0$ for $c_{k} \in \mathcal{C}_{q_{i}}$ but $c_{k} \notin \mathcal{C}_{q_{i+1}}$, this means that the user can still be satisfied without such a card type. In other words, the disappeared card is not relevant to the user's information need. Even if we have to include such a card in the list, we expect to rank it in a low position.
\end{enumerate}Considering all the five cases, we formulate the label $l^{q_{i+1}}_k $ based on the function $D(r^{q_i}_k, r^{q_{i+1}}_k)$ as below:
\begin{align*}
	D(r^{q_i}_k, r^{q_{i+1}}_k)=\left\{\begin{matrix}
		r^{q_i}_k-r^{q_{i+1}}_k; \quad c_k \in \mathcal{C}^{q_{i+1}} \wedge c_k \in \mathcal{C}^{q_i}\\ 
		d_+; \quad c_k \in \mathcal{C}^{q_{i+1}} \wedge c_k \notin \mathcal{C}^{q_i}\\ 
		d_-; \quad c_k \notin \mathcal{C}^{q_{i+1}} \wedge c_k \in \mathcal{C}^{q_i}
	\end{matrix}\right.
\end{align*}
in which $d_+$ and $d_-$ are the default values for the cards appeared or disappeared in the reformulation. We illustrate this strategy in an example similar to the one shown before, i.e., $l^{q_{2}} = +1$ and $l^{q_{1}} = -1$ as below:
\begin{align}
	\pi_{q_{1}}: c_1 \succ c_2 \succ c_3 \succ c_4 \nonumber \\
	\pi_{q_{2}}: c_3 \succ c_2 \succ c_5 \succ c_1 \nonumber
\end{align}Then, based on {\tt Movement-based Pointwise Labeling} we can derive query-card labels as shown below:
\begin{align}
	&l^{q_{2}}_{c_{1}} = -3,  l^{q_{2}}_{c_{2}} = 0, l^{q_{2}}_{c_{3}} = +2,\nonumber \\
	&l^{q_{2}}_{c_{4}} = d_-, l^{q_{2}}_{c_{5}} = d_+\nonumber
\end{align}
In our work, we empirically tested the choice of $d_+$ and $d_-$, and we found that a moderate magnitude of their value is satisfactory for the card ranking performance. As shown in Section~\ref{sec:experiment}, we choose to set $d_+=+1$ and $d_-=-1$. We shall also notice that two design choices are underlaid in our strategy. First, we chose a specific formulation of function $D$, which may also be formulated differently. Our experimental evaluation shows that such a choice results in a reasonable card ranking performance, while we leave more elaborated design of this function to future work. Second, our labeling strategy takes in viewpoint from the reformulation, as shown in the example, $\mathbf{q}_2$. We may also take the view point from $\mathbf{q}_1$, which would result in a symmetric and equivalent labeling outcome. 

\subsection{Pairwise Labeling}\label{sec:pairwise}
Strategies present in the previous sub-section focus on deriving labels for each query-card pair. Here, we discuss a labeling method to obtain pairwise preferences between two cards from the data. The pairwise labeling strategy allows us to identify the relative contributions of each individual cards through the reformulation process.
\begin{theorem}[Pairwise Labeling]
	If $l^{q} = +1$, for all cards pairs ($c_{i}$, $c_{j}$) where $c_{i} \in \mathcal{C}_{q} \wedge c_{j} \in \mathcal{C}_{q}$ and $r_{i}^{q} < r_{j}^{q}$, let $l_{c_i \succ c_j}^{q} = +1$. If $l^{q} = -1$, for all cards pairs ($c_{i}$, $c_{j}$) where $c_{i} \in \mathcal{C}_{q} \wedge c_{j} \in \mathcal{C}_{q}$ and $r_{i}^{q} < r_{j}^{q}$, let $l_{c_i \succ c_j}^{q} = -1$.
\end{theorem}where $l_{i \succ j}^{q}$ is the label for the card pair $(c_{i}, c_{j})$, meaning that whether the card $c_{i}$ is preferred over the card $c_{j}$. Using Example \ref{eq:example_1}, we can derive following labels:
\begin{align}
	l^{q_{1}}_{c_{1} \succ c_{2}} = -1, l^{q_{1}}_{c_{1} \succ c_{3}} = -1, l^{q_{1}}_{c_{1} \succ c_{4}} = -1 \nonumber \\
	l^{q_{1}}_{c_{2} \succ c_{3}} = -1, l^{q_{1}}_{c_{2} \succ c_{4}} = -1, l^{q_{1}}_{c_{3} \succ c_{4}} = -1 \nonumber \\
	l^{q_{2}}_{c_{3} \succ c_{2}} = +1, l^{q_{2}}_{c_{3} \succ c_{5}} = +1, l^{q_{2}}_{c_{3} \succ c_{4}} = +1 \nonumber \\
	l^{q_{2}}_{c_{2} \succ c_{4}} = +1, l^{q_{2}}_{c_{2} \succ c_{5}} = +1, l^{q_{2}}_{c_{5} \succ c_{4}} = +1 \nonumber
\end{align}Given $K$ cards in $\mathcal{C}_{q}$, for pointwise methods, $O(K)$ labels would be derived while pairwise methods would derive $O(K^{2})$ labels, which is significantly more.

While it is straightforward to derive pairwise preferences as above, for a trained model, it is NP-hard to obtain an optimal ranking from predicted preferences~\cite{Liu2009} although approximations do exist for rankings with less agreements with predicted preferences.

Therefore, we provide another approximation strategy to the pairwise labeling:
\begin{theorem}[Approximated Pairwise Labeling]
	By using Pairwise-Labeling, for $l_{i \succ j}^{q} = +1$, let $l_{i}^{q} = +1$ and $l_{j}^{q} = -1$. For $l_{i \succ j}^{q} = -1$, let $l_{i}^{q} = -1$ and $l_{j}^{q} = +1$.
\end{theorem}This mechanism essentially breaks down pairwise preferences to pointwise ones while keeping the relative ordering. For the query $\pi_{q_{1}}$ in Example \ref{eq:example_1}, we could have:
\begin{align}
	l_{c_{1}}^{q_{1}} = +1, l_{c_{2}}^{q_{1}} = -1, l_{c_{1}}^{q_{1}} = +1, l_{c_{3}}^{q_{1}} = -1, l_{c_{1}}^{q_{1}} = +1, l_{c_{4}}^{q_{1}} = -1 \nonumber \\
	l_{c_{2}}^{q_{1}} = +1, l_{c_{3}}^{q_{1}} = -1, l_{c_{2}}^{q_{1}} = +1, l_{c_{4}}^{q_{1}} = -1, l_{c_{3}}^{q_{1}} = +1, l_{c_{4}}^{q_{1}} = -1 \nonumber
\end{align}If we combine multiple labels for the same card into one, the labeling result yields as:
\begin{align}
	l_{c_{1}}^{q_{1}} = +3, l_{c_{2}}^{q_{1}} = +1, l_{c_{3}}^{q_{1}} = -1, l_{c_{4}}^{q_{1}} = -3 \nonumber
\end{align}It turns out that, the labeling results are similar to the one used by {\tt Discounted-Pointwise-Labeling} but symmetric emphasizing/penalizing the top/bottom results. Comparing to the true pairwise case, {\tt Approximated-Pairwise-Labeling} has a $O(K)$ scoring time for $K$ cards. Therefore, we stick to this method in later experiments.

\subsection{Listwise Labeling}\label{sec:listwise}
Apart from the pointwise and pairwise labeling strategies, we further propose a listwise labeling strategy. We shall point out that in literature listwise learning-to-rank techniques were designed in a way to optimize approximately  ranking loss, such as {\tt NDCG} or {\tt ERR}. However, labels are prerequisite requirements for those methods and they do not tackle the issue of obtaining labels. In this work, the listwise labeling strategy is substantially different from the previous work in {\tt LtR} in the sense that we focus on deriving listwise labels, instead of training ranking models. We define the listwise labeling strategy as follows:
\begin{theorem}[Listwise Labeling]
When $l^{q_i}=-1 \wedge l^{q_{i+1}}=+1$, we label $l^{\pi_{q_{i}}}=-1$, and $l^{\pi_{q_{i+1}}}=+1$.
\end{theorem}where $l^{\pi_{q}}$ represents the label for \textbf{ranking} of cards. Taking Example \ref{eq:example_1} again as an example, listwise labels are defined for as:
\begin{align}
l^{\pi_{q_{1}}} = l^{c_1 \succ c_2 \succ c_3 \succ c_4} = -1 \nonumber \\
l^{\pi_{q_{2}}} = l^{c_3 \succ c_2 \succ c_5 \succ c_4} = +1 \nonumber
\end{align}Note that in this strategy we actually label the whole {\tt QPV} rather than query-card pair.

We shall emphasize two potential limitations of the listwise strategy and our consideration in respect to them:
\begin{itemize}
\item \textbf{Feasibility} In testing, a listwise ranking model would take permutations of all possible subsets of the card set as input and choose the one with the highest predicted score. Such approach is not practical due to its $O(2^{K})$ complexity for $K$ cards. In this sense, the listwise strategy is not able to scale up for a large (or even small) number of cards. However, we consider this strategy should be applicable in practice for two reasons. First, since the card ranking model only serves to predict the relevance of a handful of relevant cards (cases where $K<=5$), the actual running time of predictions only depends on those cards. Second, for a set of relevant cards, there could be some product design constraints that pre-set positioning rules for some cards. For instance, {\tt Web} card is designed to be always placed in the bottom of the list. For this reason, the actual possible rankings are in a small number. As a result, it is feasible to run an effective listwise ranking model in production.  
\item \textbf{Generalization} As a matter of fact, by utilizing the listwise strategy, a ranking model would only capture relevance at the list level, it is then limited in its ability to generalize. Specifically, if a particular card ranking list is not observed in the training set (i.e., the set of data we use to training the ranking model), the model is impossible to predict the relevance of such a list. In other words, all the card rankings that a model can learn are limited to observed lists. However, as mentioned above, the product design has set up quite a few card ranking constraints, and thus, in practice we do not need to assess all the possible rankings of cards. We show in our experiments that the listwise strategy can allow us to train a ranking model that performs as competitive as other alternatives.
\end{itemize}
Although we demonstrated in our work that the listwise labeling strategy is practically applicable, we do acknowledge that further efforts that lead to addressing the above limitations are highly valuable. We leave it to one of our future directions.

\subsection{Learning to Label}\label{sec:l2l}
The last strategy we propose, namely, \textit{Learning to Label} ({\tt LtL}) is to exploit an additional learning algorithm for estimating the importance of each card on a query. The idea is borrowed from \textit{multi-touch attribution} (e.g., \cite{Shao2011}) in online advertising where regression models are used to allocate credits, i.e. conversions, to multiple advertising channels. Here, we want to allocate a credit, a {\tt QPV}-level label, into different cards and use those distributed credits as pseudo labels to train ranking models. We start from a simple form to decompose $l^{q_{i}}$:
\begin{align}
	l^{q_{i}} = \mbox{Logistic}(\theta_{0}^{w} + \sum_{c} \theta_{c}^{q_{i}}), \forall q_{i} == w \wedge c \in \mathcal{C}_{q_{i}}
\end{align}where $\theta_{0}^{w}$ is the query-term-level bias, representing the natural uncertainty of the query, $\theta_{c}^{q_{i}}$ is the credit for card $c$ in {\tt QPV} $\mathbf{q_{i}}$ and $\mbox{Logistic}(x) = 1/[1+\exp(-x)]$. Logistic function is used as labels are binary. Ideally, $\theta_{c}^{q_{i}}$ should be different for different {\tt QPV} $\mathbf{q}_{i}$. The central problem is that both $\theta_{0}^{w}$ and $\theta_{c}^{q_{i}}$ are unknown for all $\mathbf{q}_{i}$. If a query term $w$ has $D$ {\tt QPV}s with average $K$ cards, the problem yields $D \times K + 1$ unknown variables for $D$ equations, making the problem hard to solve.

Instead of directly tackling $\theta_{c}^{q_{i}}$, we take the following feature-based approach by obtaining $\theta_{c}^{q_{i}}$ from some simple features, resulting in much less parameters to learn:
\begin{align}\label{eq:ltl}
	\theta_{c}^{q_{i}} = x_{c,\mbox{click}} \theta_{c,\mbox{click}}^{w} + x_{c,\mbox{view}} \theta_{c,\mbox{view}}^{w}
\end{align}where $x_{c,\mbox{click}}$ is an indicator feature, representing card $c$ being clicked, $x_{c,\mbox{view}}$ is an indicator feature, representing card $c$ being viewed while $\theta_{c,\mbox{click}}^{w}$ and $\theta_{c,\mbox{view}}^{w}$ are corresponding weights. Note that, both $\theta_{c,\mbox{click}}^{w}$ and $\theta_{c,\mbox{view}}^{w}$ are the same for the query term $w$ across all {\tt QPV}s where $q_{i} == w$. Thus, we reduce all unknown parameters from $D \times K + 1$ to $2 \times K + 1$ for $D$ {\tt QPV}s. For $M$ query terms, Equation \ref{eq:ltl} indicates $M$ separate regression problems and the whole setting can be embarrassingly parallelized. Note that, more features can be used, but in this paper, for simplicity, we only use these two features.

As Equation \ref{eq:ltl} implies, $\mathbb{E}[\theta_{c}^{q_{i}}]$, the expected credit of a card $c$, can be computed as:
\begin{align}\label{eq:value}
	\mathbb{E}[\theta_{c}^{q_{i}}] = \mathbb{E}[x_{c,\mbox{click}} \theta_{c,\mbox{click}}^{w}] + \mathbb{E}[x_{c,\mbox{view}} \theta_{c,\mbox{view}}^{w}]
\end{align}where $\mathbb{E}[x_{c,\mbox{click}} \theta_{c,\mbox{click}}^{w}]$ is essentially the mean of the feature value multiplies the learned weight, similarly for $\mathbb{E}[x_{c,\mbox{view}} \theta_{c,\mbox{view}}^{w}]$. We call $\mathbb{E}[\theta_{c}^{q_{i}}]$ the ``total value'' of the card $c$, $\mathbb{E}[x_{c,\mbox{click}} \theta_{c,\mbox{click}}^{w}]$ the ``click value'' and $\mathbb{E}[x_{c,\mbox{view}} \theta_{c,\mbox{view}}^{w}]$ the ``view value''. The total value of a card can be seen as an average contribution of a card with respect to reformulation. We formalize the {\tt LtL} strategy as below:
\begin{theorem}[Learning-to-Label]
	For $\mathbf{q_{i}}$, let $l_{c_{k}}^{q^{i}} = \theta_{c_{k}}^{q_{i}}$ where $\theta_{c_{k}}^{q_{i}}$ is computed by Equation \ref{eq:ltl}.
\end{theorem}Following Example \ref{eq:example_1}, and further we assume that none of the cards was clicked in $\mathbf{q}_1$, and $c_5$ was clicked in $\mathbf{q}_2$, then, we have the labels as show below:
\begin{align}
&l^{q_{1}}_{c_{1}} = \theta_{c_{1},\mbox{view}}^{q_1},  l^{q_{1}}_{c_{2}} = \theta_{c_{2},\mbox{view}}^{q_1}, l^{q_{1}}_{c_{3}} = \theta_{c_{3},\mbox{view}}^{q_1}, \nonumber \\
&l^{q_{1}}_{c_{4}} = \theta_{c_{4},\mbox{view}}^{q_1}, l^{q_{2}}_{c_{3}} = \theta_{c_{3},\mbox{view}}^{q_2},  l^{q_{2}}_{c_{2}} = \theta_{c_{2},\mbox{view}}^{q_2}, \nonumber \\
&l^{q_{2}}_{c_{5}} = \theta_{c_{5},\mbox{view}}^{q_2}+\theta_{c_{5},\mbox{click}}^{q_2}, l^{q_{2}}_{c_{4}} =\theta_{c_{4},\mbox{view}}^{q_2}\nonumber
\end{align}

\subsection{Alternative Labeling}\label{sec:alternative}
\begin{table}
	\caption{\textbf{An example of a few judgments on ``{\tt Facebook}''}}
	\begin{center}
		\begin{tabular}{l||l}
			\textbf{Card Name} & \textbf{Judgment} \\ \hline
			{\tt NavigationCard} & {\tt Excellent} \\
			{\tt WebCard} & {\tt Good} \\
			{\tt NewsCard} & {\tt Good} \\
			{\tt FinanceCard} & {\tt Neutral} \\
			{\tt VideoCard} & {\tt Neutral} \\
			{\tt ImageCard} & {\tt Poor} \\
			{\tt LocalCard} & {\tt Very Poor} \\
			{\tt WeatherCard} & {\tt Very Poor} \\ \hline
		\end{tabular}
	\end{center}
	\label{tab:judgments}
\end{table}
Other than labeling approaches derived from query reformulations, alternative methods do exist for ranking cards. Here, we discuss two important ones.

\medskip

{\bfseries Click-Through-Rate Labeling}: If a user clicks on links shown on a card, it can be interpreted as the user is interested in the card. Thus, it is reasonable to use Click-Through-Rate ({\tt CTR}) as a signal of relevance, as similar ideas exploited before~\cite{Joachims2002,Joachims2007}. Here, {\tt CTR} is computed as the number of links from a card got clicked normalized by the total number of links shown on the card. Thus, a higher {\tt CTR} represents a higher degree of relevance of a card with respect to the query. However, as mentioned in \S\ref{sec:introduction}, not all cards contain links, like {\tt WeatherCard} and {\tt Q2ACard}. Therefore, {\tt CTR}-based labels can only drive user engagements on link-based cards. Nevertheless, this method is a strong baseline to consider. Note that, as {\tt CTR} is computed on the query-card level, it can also treated as a pointwise method.

\medskip

{\bfseries Human Judgments Labeling}: As mentioned in \S\ref{sec:related_work}, it is a standard method to utilize human judgments to train ranking models in previous research of vertical selection or federated search. The main limitation of such a approach is that, the label is not defined on {\tt QPV} but on query-card level. Thus, it looses the way to quantify the uncertainty of query reformulations on a same query and models trained on such labels tend to have strong bias towards one particular outcome (e.g., either reformulated or not-reformulated). Even though it has limitations, it still has advantages for the scenarios like launching a new product where user feedback data is not available. In this paper, we randomly sample $600$ top queries from the mobile query log of a major search engine with $15$ cards, yielding $9,000$ human judgments in the scale of \{{\tt Excellent}, {\tt Good}, {\tt Neutral}, {\tt Poor}, {\tt Very Poor}\}. One example of a few judgments is shown in Table \ref{tab:judgments}, for the query ``{\tt Facebook}''. We can observe two additional drawbacks of human judgments: 1) it lacks of a ranking of cards as different cards may have the same relevance judgments and 2) maintaining, revising and adding judgments are tremendously time-consuming. One may argue that this human label data set is small but as we would point out later, a much larger set does not solve issues of human judgments.
\section{Ranking Model}\label{sec:gbdt}
In the previous section, we mainly deal with the problem of deriving labels from a {\tt QPV}-level user engagement metric, {\tt QR}. Here, we present a state-of-the-art ranking framework to train models with those labels. Recall that each strategy present in \S\ref{sec:card_ranking} derives labels on query-card, query-card-pair or query-card-list level. For each query $q$ and the card set $\mathcal{C}$, we can construct corresponding feature vectors $\mathbf{x}_{q}$ and $\mathbf{x}_{\mathcal{C}}$. A ranking model $F(\mathbf{x}_{q}, \mathbf{x}_{\mathcal{C}}) \rightarrow \pi_{q}$ takes such feature vectors and outputs a ordered list of cards. In theory, for each query, $F$ would evaluate all possible candidates of cards, pairs of cards or list of cards. In practice, this is never the case given that for each query, only a handful of cards could be relevant and therefore, the final output is almost always a subset of cards while other cards are decided not shown to the user. Note that, the step of deciding relevant cards can be done by the ranking model but usually is done through a simpler function with taking less features. It is out of the scope of this paper to describe such a function. Basically, one can assume that, a pool of small number of cards would be present to $F$ after this relevance evaluation step for each query. For $F$, it has following scenarios:
\begin{enumerate}[noitemsep,nolistsep]
	\item {\bf Pointwise Labels}: For a candidate set of cards, $F$ evaluates each card and outputs a score. The final $\pi$ is obtained by sorting.
	\item {\bf Pairwise Labels}: As mentioned in \S\ref{sec:pairwise}, evaluating all possible pairs of cards and obtain the optimal ranking is NP-hard. For {\tt Approximated-Pairwise-Labeling}, it essentially has the same procedure to obtain a ranking as pointwise methods.
	\item {\bf Listwise Labels}: As mentioned in \S\ref{sec:listwise}, in theory, $F$ would evaluate all possible rankings but in practice, $F$ evaluates on rankings that haven been shown to users in the past.
	\item {\bf Learning To Label}: It has the exact same procedure as pointwise ones.
\end{enumerate}For these scenarios, feature sets are adapted to meet their criterion. Note that, depending on the value of labels, different strategies would yield either classification problems or regression problems.

In this paper, we use {\tt GBT} algorithm~\cite{Friedman2001} to learn $F$ for all scenarios mentioned above. {\tt GBT} is an additive regression algorithm consisting of an ensemble of trees, fitted to current residuals, gradients of the loss function, in a forward step-wise manner. It iteratively fits an additive model as:
\begin{align*}
	f_{K}(x) = T_{0}(x; \Theta) + \lambda \sum_{k=1}^{K} \beta_{k} T_{k}(x;\Theta_{k})
\end{align*}such that a certain loss function $L(y_{i}, f_{K}(x_{i}))$ (e.g., square loss, logistic loss) is minimized, where $T_{k}(x; \Theta_{k})$ is a tree at iteration $k$, weighted by a parameter $\beta_{k}$, with a finite number of parameters $\Theta_{k}$, and $\lambda$ is the learning rate. At iteration $k$, tree $T_{k}(x; \beta)$ is induced to fit the negative gradient by least squares. That is:
\begin{align*}
	\hat{\Theta} = \arg\min_{\beta} \sum_{i}^{N} w_{i}(-G_{ik} - \beta_{k}T_{k}(x_{i}); \Theta)^{2}
\end{align*}where $w_{i}$ is the weight for data instance $i$, which is usually set to $1$, and $G_{it}$ is the gradient over the current prediction function: $G_{ik} = \Bigr[ \frac{\partial L(y_{i}, f(x_{i}))}{\partial f(x_{i})} \Bigl]_{f = f_{k-1}}$. The optimal weights of tree $\beta_{k}$ are determined by $
\beta_{k} = \arg\min_{\beta} \sum_{i}^{N} L(y_{i}, f_{k-1}(x_{i}) + \beta T(x_{i}, \theta))$. More details about {\tt GBT}, please refer to \cite{Friedman2001,Zheng2008}.

\section{Experimental Evaluation}\label{sec:experiment}
To evaluate the effectiveness of our proposed methods, we use a sample of {\tt QPV} data from a major search engine. In particular, the data is randomly sampled from two weeks' data produced by a production mobile card ranking system. It contains $4,154,958$ distinct queries and $7,327,012$ {\tt QPV}s.

\medskip

{\bfseries Comparisons}: We compare the following approaches of labeling in this section: 1) {\bfseries Pointwise Labeling}, mentioned in \S\ref{sec:pointwise},  includes {\tt Na\"{i}vely-Pointwise-Labeling} ({\tt NPL}), {\tt Discounted-Pointwise-Labeling} ({\tt DPL}) and {\tt Movement-based-Pointwise-Labeling} ({\tt MPL}), 2) {\bfseries Pairwise Labeling}, {\tt Approximated-Pairwise-Labeling}, mentioned in \S\ref{sec:pairwise}), abbreviated as {\tt APL} 3) {\bfseries Listwise Labeling}, mentioned in \S\ref{sec:listwise}, abbreviated as {\tt LL}, 4) {\bfseries Learning To Label}, mentioned in \S\ref{sec:l2l}, abbreviated as {\tt LtL}, 5) {\bfseries Click-Through-Rate Lableing}, mentioned in \S\ref{sec:alternative}, abbreviated as {\tt CTR}, and 6) {\bfseries Human Judgment Lableing}, mentioned in \S\ref{sec:alternative}, abbreviated as {\tt Human}.

\medskip
\begin{figure}[t]
	\centering
	\includegraphics[scale=0.32]{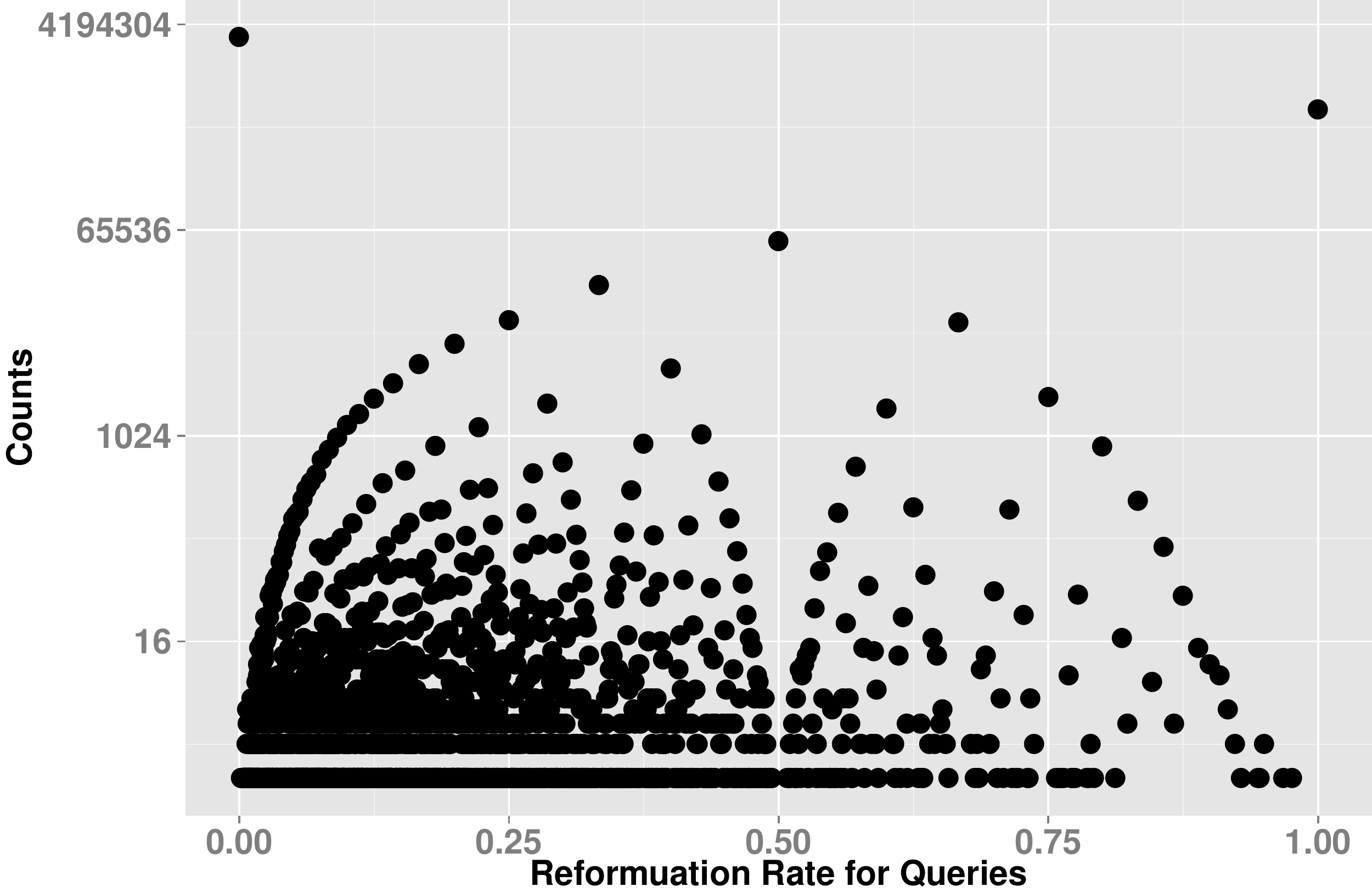}
	\caption{\textbf{Counts of the ratio of reformulated {\tt QPV}s per query} (Y-axis is log-scaled).}
	\label{fig:dist_rr}
\end{figure}

{\bfseries Evaluation Protocol}: The key difficulty of training card ranking models is that, there is no established method to properly evaluate them. As mentioned in \S\ref{sec:related_work}, the classic evaluation method used in the problem of vertical selection or federated search requires human judgments as ground-truth labels and traditional ranking metrics such as {\tt NDCG}, {\tt MAP} and {\tt ERR} are used to compare different models. Here, we use a different approach. For each {\tt QPV}, we use $l^{q}$ as the ground-truth label, which is either $+1$ or $-1$ for the whole $\pi_{q}$. When testing, a predicted ranked list $\hat{\pi}_{q}$ is produced by a ranking model. We compute two metrics:
\begin{itemize}
	\item {\bf True-Positive-Ratio} ({\tt TPR}) is defined as:
	\begin{align}
		\frac{1}{N^{+}}\sum_{i=1}^{N} \mathbb{I}(\hat{\pi}_{q_{i}} == \pi_{q_{i}}) \times \mathbb{I}(l^{p_{i}} == +1) \nonumber
	\end{align}
	where $N^{+} = \sum_{i=1}^{N}\mathbb{I}(l^{p_{i}} == +1)$, the total number of positive ranked lists and $\mathbb{I}(x) = 1$ when $x$ is true otherwise $0$.
	\item {\bf True-Negative-Ratio} ({\tt TNR}) is defined as:
	\begin{align}
		\frac{1}{N^{-}}\sum_{i=1}^{N} \mathbb{I}(\hat{\pi}_{q_{i}} == \pi_{q_{i}}) \times \mathbb{I}(l^{p_{i}} == -1) \nonumber
	\end{align}
	where $N^{-} = \sum_{i=1}^{N}\mathbb{I}(l^{p_{i}} == -1)$, the total number of negative ranked lists.
\end{itemize}Note that, for both metrics, we require $\hat{\pi}_{q}$ matches $\pi_{q}$ \textbf{exactly}. Both $\mbox{TPR}$ and $\mbox{TNR}$ resemble the importance sampling technique used in offline A/B testing evaluation methods~\cite{Li2015a}. As $\mbox{TPR}$ emphasizes that a ranker could match non-reformulated ranked lists, which is a positive sign of a model and $\mbox{TNR}$ emphasizes that a ranker could match reformulated cases, which is a negative sign of a model. Therefore, a good ranker is the one that has a high score in $\mbox{TPR}$ but achieves a low one in $\mbox{TNR}$. Given this observation, we define a $F$-Measure as:
\begin{align}
	\frac{2 \times \mbox{TPR} \times (1- \mbox{TNR})}{\mbox{TPR} + (1- \mbox{TNR})} \nonumber
\end{align}which defines as a \textit{harmonic mean} between $\mbox{TPR}$ and $1-\mbox{TNR}$ where $1- \mbox{TNR}$ is used to penalize a ranker which matches negative ranked lists. Under this definition, the ranker that produces the dataset has $\mbox{TPR}=100\%$ and $\mbox{TNR}=100\%$, resulting in $0.0$ in $F$-Measure.

\medskip

{\bfseries Parameters}: We chose parameters based on $5$-cross validations and results are reported from cross-validation as well and they are statistical significant. For {\tt GBT}, we choose $67$ trees, $10$ nodes and $0.1$ shrinkage from cross validation and fix them for all models. Squared-loss is used in {\tt GBT} and we found that logistic-loss does not give any significant difference in terms of evaluation metrics introduced above. For more discussion about {\tt GBT}, please refer to \cite{Friedman2001,Zheng2008} for the details about these parameters.

\medskip

{\bfseries Features}: As mentioned in \S\ref{sec:gbdt}, we use a number of feature groups, resulting in approximately $1,000$ in total. In particular, we have:
\begin{itemize}
	\item {\bfseries Lexical Features}: These features include unigram, bigram and language models of query terms, which have been hashed \cite{Weinberger2009} into a fixed number of bins. Then, a simpler linear model is trained through these lexical features to indicate whether a card is relevant to a query.
	\item {\bfseries Query Intent Features}: Queries are classified into a $200$ hierarchical taxonomy where each node represents an intent. Certain intent might have strong indication for a particular card. For instance, a \textit{local intent} may imply {\tt LocalCard} or {\tt WeatherCard} stronger than other intents.
	\item {\bfseries Card Backend Features}: For a given query, whether a card's backend system handles it is also important factor for a ranking system to consider. For instance, even if a query has a local intent, the {\tt LocalCard} may not find relevant stores, restaurants and other local business. Therefore, the card ranking system would incorporate the returned results or relevance scores from a backend as signals to leverage.
	\item {\bfseries Click Feedback Features}: Sometimes, a relevance between a query and a card may be influenced by some temporal factors. For instance, in general, the query ``{\tt Apple}'' may not have a news intent. But, if the Apple Company announces new product releases, the {\tt NewsCard} would become relevant during that period of time. Thus, we have features to track click-through-rate of URLs related to a query and relate them to a card, capturing the temporal dynamics of a card with respect to certain queries.
\end{itemize}As this paper is not about card ranking models, we do not discuss features in detail.

\subsection{Basic Statistics}\label{sec:basic}
\begin{figure}
	\centering
	\includegraphics[scale=0.35]{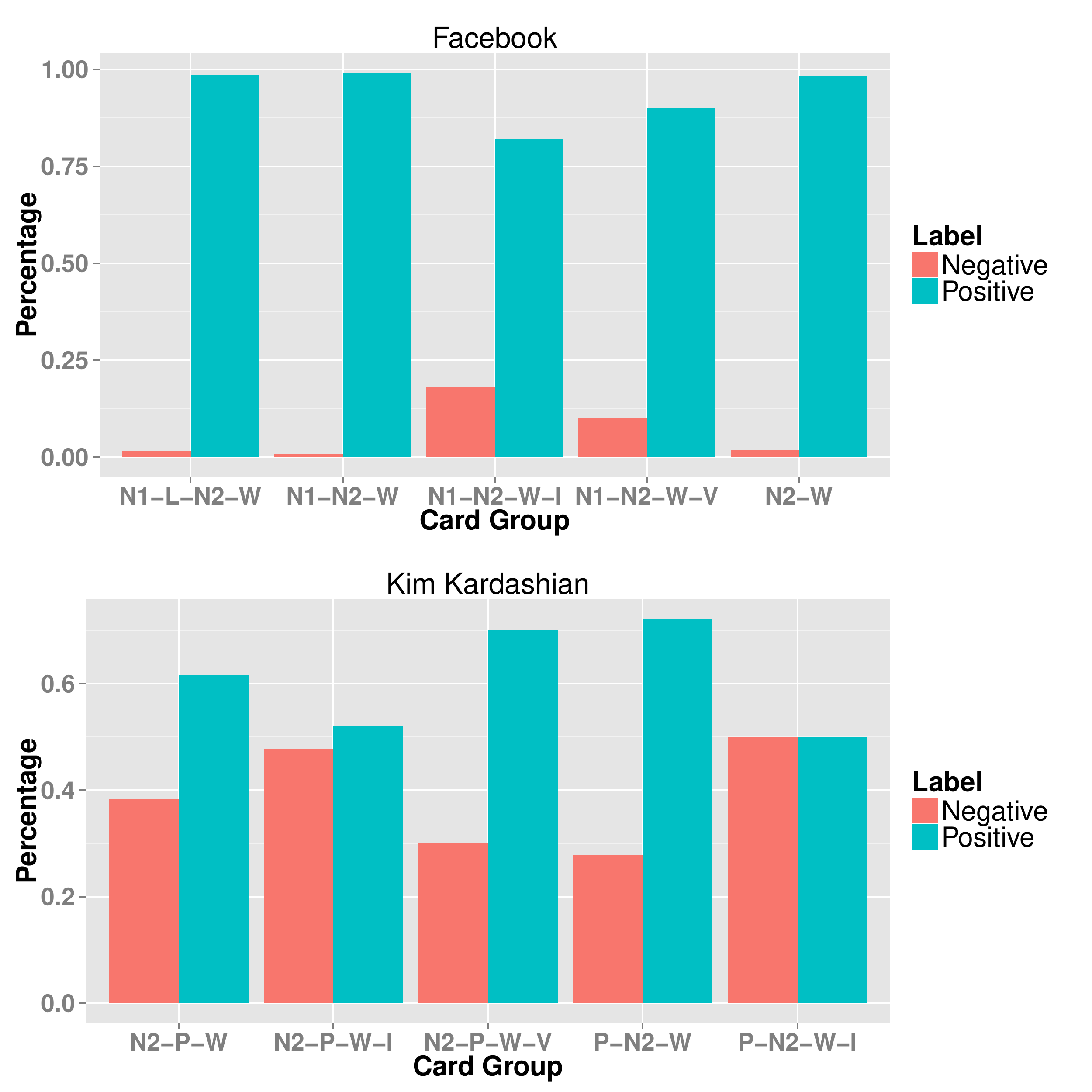}
	\caption{\textbf{Percentage of positive and negative examples per card group for queries ``{\tt Facebook}'' (top) and ``{\tt Kim Kardashin}'' where X-axis shows different card groups and Y-axis shows the percentage. {\tt N1}, {\tt N2}, {\tt L}, {\tt W}, {\tt I}, {\tt V} and {\tt P} represent {\tt NavigationCard}, {\tt NewsCard}, {\tt LocalCard}, {\tt WebCard}, {\tt ImageCard}, {\tt VideoCard} and {\tt PersonCard}.}}
	\label{fig:facebook_kim}
\end{figure}
\begin{table}[t]
	\caption{\textbf{The percentage of the number of cards}}
	\begin{center}
		\begin{tabular}{c||r|r|r}
			\textbf{\# Cards} & Per-\textbf{{\tt QPV}} & Per-Query & Per-Label(+/-)\\ \hline
			2 & 69.5228\% & 89.5710\% & 84.30\%/15.70\% \\
			3 & 29.0713\% & 9.2012\% & 92.62\%/7.38\%\\
			4 & 1.3854\% & 1.0068\% & 92.55\%/7.45\%\\
			5 & 0.0170\% & 0.1681\% & 91.49\%/8.51\%\\
		\end{tabular}
	\end{center}
	\label{tab:pvid_per_card}
\end{table}
In this sub-section, we firstly show some characteristics of the dataset. First, starting from Figure \ref{fig:dist_rr}, it shows the number of queries with certain ratio of reformulated {\tt QPV}s. The point on the upper left corner is the number of queries with $0$ ratio of reformulated {\tt QPV}s, meaning that all {\tt QPV}s have label $1$ while the upper right corner represents queries with $1$ ratio of reformulated {\tt QPV}s. After reviewing the data, we found that both points are from queries with little {\tt QPV}s, demonstrating extreme cases. Other parts of the figure do not reveal strong patterns, which might indicate that all queries possibly can be reformulated no matter they are top queries or long-tail ones.

We take two queries ``{\tt Facebook}'' (shown at the top) and ``{\tt Kim Kardashian}'' (shown at the bottom) as examples, shown in Figure \ref{fig:facebook_kim}. We show several card groups for each query with their corresponding percentage of positive {\tt QPV}s and negative ones. A card group is defined as a $\pi$ of cards for the query. As we can see, no matter which card group, for both queries, there exists reformulated {\tt QPV}s, even though some card groups (e.g, {\tt N1-L-N2-W} and {\tt N1-N2-W} and {\tt N2-W}) have a very low reformulation ratio. Additionally, there is no one single card group with high ratio of positive labels. In other words, a card group with less reformulation ratio does \textbf{not} imply that other card groups are less likely to be positive. From these two examples, we can see that, some queries are inherently more likely to be reformulated than others.

In Table \ref{tab:pvid_per_card}, we show the percentage of {\tt QPV}s containing $K$ of cards where $K \ge 2$ in the second column of the table. We can see that nearly $70\%$ of {\tt QPV}s which contains $2$ cards while almost no {\tt QPV}s show $5$ more cards and beyond. This distribution is very intuitive as mobile devices have very constrained screens. The third column shows the percentage of queries which contain a certain number of cards, demonstrating that a large portion of queries have $2-3$ cards. The last column of the table shows that, for a given number of cards, what percentage of {\tt QPV}s has a positive label or not. Note that, the ratio of positive versus negative labels is highly skewed as negative ones are sparse across different $K$, indicating that, learning to avoid negative examples is an inherent difficult task.

\subsection{Comparisons of Labeling Strategies}\label{sec:comparison}
\begin{table*}[t]
	\caption{\textbf{The comparison of different strategies. Results are statistical significance, sorted by $F$-Measure.}}
	\begin{center}
		\begin{tabular}{l||llll}
			\textbf{Method Name} & \textbf{True-Positive-Ratio} & \textbf{True-Negative-Ratio} & 1-\textbf{True-Negative-Ratio} & $F$-\textbf{Measure} \\ \hline
			{\tt LtL} & 74.40\% & 40.83\% & 59.17\% & 65.91\% \\
			{\tt APL} & 53.31\% & 33.42\% & 66.58\% & 59.21\% \\
			{\tt DPL} & 69.44\%	& 54.87\% & 45.13\% & 54.71\% \\
			{\tt NPL} & 64.50\% & 57.84\% & 42.16\% & 50.99\% \\
			{\tt MPL} & 53.58\% & 63.09\% & 36.91\% & 43.71\% \\
			{\tt CTR} & 26.39\% & 24.69\% & 75.31\% & 39.08\% \\
			{\tt Human} & 12.96\% & 17.47\% & 82.53\% & 22.40\% \\ \hline
		\end{tabular}
	\end{center}
	\label{tab:comparison}
\end{table*}
\begin{table*}[t]
	\caption{\textbf{Two examples of values and parameters of card contributions learned from {\tt LtL}. The top part of the table shows results for ``{\tt Barack Obama}'' while the bottom part is for ``{\tt Apple}''. ``C.'' means ``Click'' and ``V.'' represents ``View''.}}
	\begin{center}
		\begin{tabular}{l||cccccc||c}
			\textbf{Card Name} & \textbf{C. Value} & \textbf{C. Weight} & \textbf{C. Mean} & \textbf{V. Value} & \textbf{V. Weight} & \textbf{V. Mean} & \textbf{Total Value} \\ \hline
			{\tt NewsCard} & 0.1261 & 2.1703 & 0.0581 & 0.1898 & 0.2229 & 0.8514 & 0.3159 \\
			{\tt WebCard} & 0.3289 & 2.1859 & 0.1505 & -0.1141 & -0.1417 & 0.8048 & 0.2148 \\
			{\tt PersonCard} & 0.3112 & 2.0582 & 0.1512 & -0.1532 & -0.1901 & 0.8060 & 0.1580 \\
			{\tt ImageCard} & 0.1417 & 1.5042 & 0.0942 & 0.0002 & 0.0013 & 0.1214 & 0.1419 \\
			{\tt VideoCard} & 0.0024 & 3.2064 & 0.0007 & -0.0006 & -0.1306 & 0.0045 & 0.0018 \\ \hline \hline
			{\tt NavigationCard} & 0.7460 & 1.5485 & 0.4817 & 0.2160 & 0.2284 & 0.9458 & 0.9620 \\
			{\tt NewsCard} & 0.0110 & 1.0210 & 0.0108 & 0.3866 & 0.4138 & 0.9342 & 0.3976 \\
			{\tt LocalCard} & 0.0126 & 1.3812 & 0.0091 & 0.0366 & 0.0533 & 0.6880 & 0.0492 \\
			{\tt ImageCard} & 0.0292 & 1.1571 & 0.0252 & 0.0068 & 0.1926 & 0.0353 & 0.0360 \\
			{\tt WebCard} & 0.0017 & 1.1054 & 0.0015 & 0.0083 & 0.1351 & 0.0615 & 0.0100 \\
			{\tt VideoCard} & 0.0013 & 1.5319 & 0.0009 & 0.0021 & 1.0835 & 0.0019 & 0.0034 \\ \hline
			
		\end{tabular}
	\end{center}
	\label{tab:weights}
\end{table*}
In this sub-section, we compare different labeling strategies in terms of $\mbox{TPR}$, $\mbox{TNR}$ and $F$-Measure, shown in Table \ref{tab:comparison}. The first phenomenal observation is that, all labeling strategies are significantly better than {\tt Human}, not only in terms of $F$-Measure but also on $\mbox{TPR}$. Human editors cannot predict what users want and indeed, the rankings induced from their assessed relevance labels do not match what users like. In addition, the ranking model for {\tt Human} does not scale, only trained from a small amount of judgments which yield in sub-optimal results while other strategies can handle millions of {\tt QPV}s.

Although {\tt CTR} is much better than {\tt Human}, it is still the second worst method in the result, which is understandable in some sense. First of all, {\tt CTR} and {\tt QR} are two different objectives. They might be related but certainly still not the same. As mentioned in \S\ref{sec:alternative}, cards may not have links. Meanwhile, it is also hard to argue that the normalized click-through-rate for different cards is a good indicator for a better ranking. For instance, even if two cards have the same number of links (e.g., say $5$), it is not always true that the one with $2$ clicks is better than the one with $1$ click. However, the result here does \textbf{not} mean that, clicks are not informative at all.

For pointwise methods, {\tt NPL}, {\tt DPL} and {\tt MPL} all perform significantly better than the baseline and {\tt CTR}. In particular, {\tt NPL} performs surprisingly well, given its simplicity. One possible explanation is that, {\tt NPL}, to some degree, resembles the idea of {\tt LL}. For example, if $\pi_{q_{i}} = \{c_{7} \succ c_{8} \succ c_{9} \}$ and $l^{q_{i}} = +1$, {\tt LL} would generate a label $l^{c_{7} \succ c_{8} \succ c_{9}} = +1$ and {\tt NPL} would generate labels: $l^{q_{i}}_{c_{7}} = +1$, $l^{q_{i}}_{c_{8}} = +1$ and $l^{q_{i}}_{c_{9}} = +1$. Essentially, two methods generate the same labels for this data instance. Although {\tt NPL} and {\tt LL} produce similar labels for many cases, they differ in some subtle scenarios. For instance, if if $\pi_{q_{j}} = \{c_{9} \succ c_{8} \succ c_{7} \}$ and $l^{q_{j}} = -1$, {\tt LL} would generate a label $l^{c_{9} \succ c_{8} \succ c_{7}} = -1$, which has nothing to do with the other label generated above. However, {\tt NPL} would generate $l^{q_{i}}_{c_{7}} = -1$, $l^{q_{i}}_{c_{8}} = -1$ and $l^{q_{i}}_{c_{9}} = -1$, which obviously interfere with the labels generated above. Therefore, we can see that, without ordering {\tt NPL} would confuse itself with the effect of different original labels of the \textbf{same} set of cards. This is also observable in terms of performance as {\tt LL} is superior to {\tt NPL}. The relative worse performance of {\tt MPL} might indicate that, the movement of a single card is overly penalized or encouraged and induced labels do not keep the order of rankings. On the other hand, {\tt DPL} performs quite well. Although it is a pointwise method, it carries over the ordering information of positive examples and negative ones and therefore, it has a high $\mbox{TPR}$, achieving a good $F$-Measure.

For pairwise methods and listwise methods, {\tt APL} achieves the second best performance. It has a relative high $\mbox{TPR}$ and low $\mbox{TNR}$, demonstrating that it has a balance between maintaining good pairs of cards and avoiding bad pairs. As mentioned before, {\tt APL} has the same prediction complexity as pointwise methods and therefore, it is a even more preferred one in terms of performance and simplicity. For {\tt LL}, as expected, it has the highest $\mbox{TPR}$, as it literally remembers good and bad cases, achieves quite good performance overall.

For {\tt LtL}, it outperforms all other methods as it has a high $\mbox{TPR}$ and a relatively low $\mbox{TNR}$. The main advantage of this method is that, it learns contributions of a card with respect to reformulations or not in a principled way. In addition, it only generates the same amount of training instances as pointwise methods, while {\tt APL}, although it has a strong performance, generates significantly more data instances as shown in examples in \S\ref{sec:pairwise}. The only obvious drawback of {\tt LtL} is that, a model needs to be trained first to obtain labels. But this shortcoming can be mitigated as the model might be trained from a large corpus and keep it constant for a while and card ranking models can be re-trained more frequently. In order to demonstrate the effectiveness of {\tt LtL}, we show regression weights learned by {\tt LtL} for two queries ``{\tt Barack Obama}''and ``{\tt Apple}'' in Table \ref{tab:weights}. The first column is the card name and columns $2-4$ represent ``click value'', click weight (learned from the model) and click mean (computed from the training set), similarly for ``view'' (columns $5-7$). The ``total value'' in the last column is defined in Equation \ref{eq:value}. Cards are sorted by ``total value'' in the table. As we can see that, ``total value'' gives a very intuitive functional explanation of cards and their engagement contributions. For instance, {\tt NewsCard} and {\tt WebCard} are comparatively much important than other cards for ``{\tt Barack Obama}'' while {\tt NavigationCard} is way more critical for ``{\tt Apple}'' as most people wanted to use {\tt NavigationCard} to quickly jump to Apple's homepage. In addition to its superior performance, {\tt LtL} can provide valuable insights that other approaches cannot offer.
 
\section{Conclusions}\label{sec:conclusion}
We have presented in this paper a comprehensive series of strategies of exploiting the users' query reformulation activities for labeling query-card relevance, based on which effective card ranking models can be optimized for mobile search. 
We demonstrated that the proposed labeling strategies achieve substantial improvement over the conventional human-judgment-based labeling strategy. In addition, our experimental results show that by directly exploiting user feedback from query reformulation we can attain a better card ranking model, compared to the conventional user feedback from CTR. Finally, the learning-to-label strategy succeeds in building discriminative query-level labeling models, which leads to a card ranking model that performs superior to other alternatives. For future work, we would explore possibilities to derive labels based on task-level search metrics  and develop ranking models to optimize them.
 

\bibliographystyle{abbrv}
\bibliography{source}
\end{document}